\documentstyle[seceq,supplement,epsf]{ptptex}

\def\nle{\ \raise.3ex\hbox{$<$}\kern-0.8em\lower.7ex\hbox{$\sim$}\ }
\def\nge{\ \raise.3ex\hbox{$>$}\kern-0.8em\lower.7ex\hbox{$\sim$}\ }

\title{
Dilution Effects on Two-Dimensional Heisenberg Antiferromagnets with
Non-Magnetic Spin-Gapped Ground State
}

\author{
Chitoshi {\sc Yasuda},$^{1,}$\footnote{E-mail address:
cyasuda@issp.u-tokyo.ac.jp} Synge {\sc Todo},$^{1,2}$ Munehisa {\sc
Matsumoto},$^1$ \\ and Hajime {\sc Takayama}$^1$
}

\inst{
$^1$Institute for Solid State Physics, University of Tokyo, Kashiwa
277-8581, Japan \\
$^2$Theoretische Physik, Eidgen\"{o}ssische Technische Hochschule,
CH-8093 Z\"{u}rich, Switzerland
}

\abst{
Dilution effects on spin-1/2 quantum Heisenberg antiferromagnets
with a non-magnetic spin-gapped ground state are
studied by means of the qunatum Monte Carlo simulation. In the site-diluted
system, an antiferromagnetic long-range order (AF-LRO) is induced at an
infinitesimal concentration of dilution due to an effective coupling
$\tilde{J}_{mn}$ between induced magnetic moments. In the bond-diluted
case, on the other hand, the AF-LRO is not induced up to a certain
concentration of dilution due to singlet pairs reformed by an AF
coupling $\tilde{J}_{\rm af}$ between the edge spins of the diluted bond
through the two-dimensional shortest
paths. The competition between $\tilde{J}_{mn}$ and $\tilde{J}_{\rm af}$
yields peculiar phenomena in the bond-diluted system.
}

\begin{document}

\maketitle

\section{Introduction}
Dilution effects on low-dimensional antiferromagnets with a spin-gapped
ground state exhibit various interesting phenomena inherent to
low-dimensional quantum spin systems. In the first inorganic
spin-Peierls compound CuGeO$_3$, the dimerized state is realized at low
temperature and the system has the spin-gapped ground
state.~\cite{hase} When a small amount of the non-magnetic impurity Zn
or Mg is substituted for Cu, the spin-gapped ground state of the pure
system changes to the antiferromagnetic (AF) long-range ordered
state,~\cite{martin,masuda,manabe} thereby the lattice dimerization is
preserved. The impurity-induced AF long-range order (LRO) has been observed
also for other spin-gapped materials such as the two-leg ladder
SrCu$_2$O$_3$~\cite{azuma} and the Haldane compound
PbNi$_2$V$_2$O$_8$.~\cite{uchiyama} The non-magnetic-impurity-induced
AF-LRO has been numerically investigated by simulating the site-diluted AF
Heisenberg model with the spin-gapped ground
state.~\cite{imada,wessel,yasuda} In the site-diluted system, the AF-LRO
is induced even at 1 \% of site dilution for the AF Heisenberg
model of the coupled alternating chains,~\cite{yasuda} suggesting the
critical concentration of non-magnetic impurities for AF-LRO to come up
is actually 0 (see also Fig.~1 below). When spins are
removed from sites, the {\it effective spins} are induced, whose
magnitudes are peaked at the neighboring site connected to the removed
site by a strong bond. They
have the same extent as the correlation length of the pure
system. Furthermore, there exists an effective coupling $\tilde{J}_{mn} \propto
(-1)^{|r_m-r_n+1|}{\rm exp}[-l/\sqrt{\xi_{\rm p}^{x}\xi_{\rm
p}^{y}}]$ between the effective spins,~\cite{nagaosa,iino} which
preserves the staggerdness with respect to the original lattice and
decays exponentially as the distance between the effective spins,
$l=|r_m-r_n|$, increases. Here $\xi_{\rm p}^{x,y}$ are the anisotropic
correlation lengths of the pure system in the two directions on a square
lattice. Thus the AF-LRO is induced at an infinitesimal concentration of
dilution for the two-dimensional (2D) site-diluted system. 

In the present work we have investigated ground-state phase transitions in
the bond-diluted 2D spin-1/2 AF Heisenberg model of the coupled alternating
chains on a square lattice by means of the quantum Monte Carlo (QMC)
simulation with the continuous-imaginary-time loop algorithm. In the
bond-diluted case, in contrast to the site-diluted one, a finite critical
concentration of dilution, $x_{\rm c}$, exists, above which the system
has the AF-LRO. In the extremely-low-concentration region, the system
has a spin gap independent of the concentration of the diluted bond,
$x$. The spin gap $\Delta$ is proportional to the squared interchain
coupling $J'^2$ significantly smaller than the original spin gap in the
pure system, $\Delta_{\rm p}$. This spin gap $\Delta$ is caused by
singlet pairs of two edge spins on the diluted bond by the AF coupling
$\tilde{J}_{\rm af}$ through the 2D shortest paths. Furthermore, when
$x$ increases, $\Delta$ starts to decrease, and tends to vanish at $x$
which is smaller than $x_{\rm c}$. This may suggest the existence of the
quantum Griffiths phase in this system.

\section{Model and method}

\begin{figure}[b]
\begin{center}
 \leavevmode
 \epsfxsize=0.6\textwidth
 \epsfbox{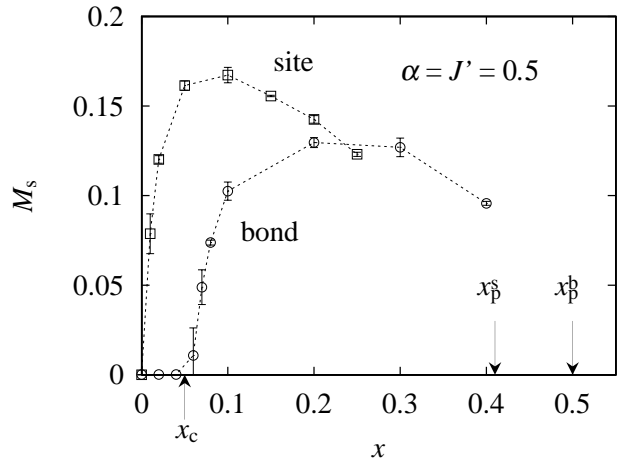}
 \caption{Concentration dependences of the staggered magnetization in the
 site-~\cite{yasuda} and bond-diluted systems with
 $\alpha=J'=0.5$. In the bond-diluted case there is the finite critical
 concentration $x_{\rm c} \simeq 0.05$. The percolation thresholds on
 the site and bond processes are denoted by $x_{\rm p}^{\rm s}$ and
 $x_{\rm p}^{\rm b}$, respectively. All the lines are guides to eyes.}
\end{center}
\vspace*{-1ex}
\end{figure}

The bond-diluted AF Heisenberg model of the coupled alternating
chains on a square lattice investigated in the present work is described
by the Hamiltonian
\begin{eqnarray}
   \label{ham}
   H &=&\sum_{i,j}\epsilon_{(2i,j)(2i+1,j)}
             {\bf S}_{2i,j}\cdot{\bf S}_{2i+1,j}
    +\alpha\sum_{i,j}\epsilon_{(2i+1,j)(2i+2,j)}
             {\bf S}_{2i+1,j}\cdot{\bf S}_{2i+2,j}  \nonumber \\
    &+&J'\sum_{i,j}\epsilon_{(i,j)(i,j+1)}
             {\bf S}_{i,j}\cdot{\bf S}_{i,j+1} \ ,
\end{eqnarray}
where 1 and $\alpha$ ($0 \le \alpha \le 1$) are the AF intrachain alternating coupling
constants, $J'$ ($\ge 0$) the AF interchain coupling constant, and ${\bf
S}_{i,j}$ the quantum spin operator with $S=1/2$ at site
($i,j$). Randomly quenched bond occupation factors
\{$\epsilon_{(i,j)(k,l)}$\} independently take either 1 or 0 with
probability $1-x$ and $x$, respectively.
The pure system described by (\ref{ham}) with
$\epsilon_{(i,j)(k,l)}=1$ for all bonds is in either a spin-gapped phase
or an AF long-range ordered phase at zero temperature depending on the strengths of $\alpha$ and $J'$.~\cite{matsumoto} In the $\alpha=0.5$ system which we
examine in the present work, there exists the spin-gapped phase for
$J'<J'_{\rm c}\simeq 0.55$.

The QMC simulations with the continuous-imaginary-time loop
algorithm~\cite{evertz,beard,todo} are carried out on $L\times L$ square
lattices with the periodic boundary conditions. For each sample with a
bond-diluted configuration, $10^3 \sim 10^4$ Monte Carlo steps (MCS) are
spent for measurement after $500 \sim 10^3$ MCS for
thermalization. Sample average for dilution is taken over $10 \sim 10^3$
samples depending on $L$, $x$, and the temperature $T$.

\section{Results and discussions}

The staggered magnetization $M_{\rm s}(x)$ at $T=0$ is evaluated by
\begin{equation}
  M_{\rm s}^{2}(x) = \lim_{L\to \infty}\lim_{T\to 0} \frac{3}{L^4}
  \sum_{i,j}(-1)^{|i-j|} \langle S_{i}^{z}S_{j}^{z}\rangle \ ,
\end{equation}
where the bracket $\langle \cdots \rangle$ denotes both the
thermal and random averages.
The concentration dependences of $M_{\rm s}$ in the site-~\cite{yasuda}
and bond-diluted systems with $\alpha=J'=0.5$ are shown in Fig. 1. In
the site-diluted case, the value of $M_{\rm s}$ becomes finite even at 1
\% of site dilution. This is caused by the strong correlation between
the effective spins due to the effective coupling $\tilde{J}_{mn}$
mentioned in $\S$ 1. In the bond-diluted case, on the other hand, there is the
finite critical concentration of $x_{\rm c}\simeq 0.05$ even for
$\alpha=J'=0.5$, which is the system near the critical point $J'_{\rm
c}\simeq 0.55$ for $\alpha=0.5$ and $x=0$. Further dilution would yield
the disappearance of $M_{\rm s}$ near the percolation
thresholds.~\cite{kato,sandvik}

Next we examine the spin gap $\Delta$ between the ground state and the first
excited state to reveal the nature of the disordered phase below $x_{\rm
c}$ in the bond-diluted system. We adopt the second-moment
method,~\cite{todo} by which the value of $\Delta^{(2)}$ defined as
\begin{equation}
  \label{delta2}
  (\Delta^{(2)})^{-1}=\lim_{L\to \infty}\lim_{\beta\to \infty}\frac{\beta}{2\pi} \sqrt{\frac{S(\pi,\pi,0)}{S(\pi,\pi,2\pi/\beta)}-1}
\ ,
\end{equation}
is generally considered to be a good estimate of $\Delta$. In
Eq. (\ref{delta2}) $\beta=1/T$ and $S(\pi,\pi,\omega)$ is the dynamic structure
factor at momentum ($\pi$, $\pi$) described by $S(\pi, \pi,
\omega)=\int_{0}^{\beta} C(\tau) {\rm e}^{i\omega\tau} {\rm d}\tau$,
where $C(\tau)$ is the correlation function of the staggered
magnetization in the imaginary-time direction:
\begin{equation}
 \label{dyna}
  C(\tau) = \frac{1}{L^2 \beta} \sum_{i,j}\int_{0}^{\beta} {\rm d}\tau' (-1)^{|i-j|}\langle S_{i}^{z}(\tau')S_{j}^{z}(\tau'+\tau) \rangle  \
\end{equation}
with $S_{i}^{z}(\tau)$ being the $z$ component of the spin operator at the
imaginary time $\tau$. Its $x$-dependence for the system with $\alpha=0.5$ and
$J'=0.3$ is shown in Fig. 3 below. For a small $x\ (\nle 0.01)$ $\Delta^{(2)}$
rapidly decreases with increasing $x$. However as we will argue below
this is an artifact of the second-moment method except for exactly at $x=0$.

\begin{figure}[t]
\begin{center}
 \leavevmode
 \epsfxsize=0.6\textwidth
 \epsfbox{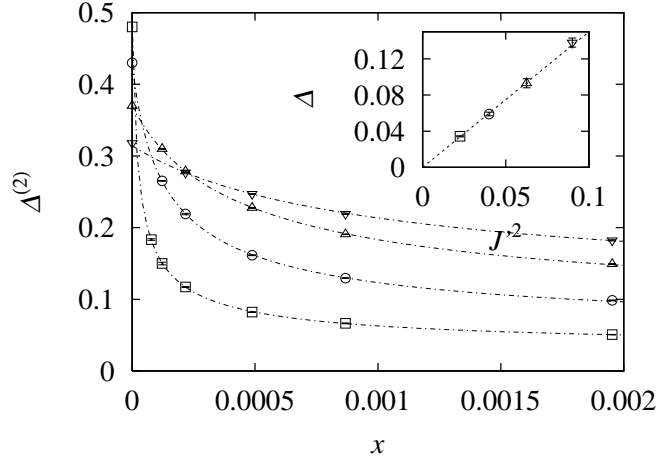}
 \caption{Dependences of $\Delta^{(2)}$ on $x$ ($=1/2L^2$) for $\alpha=0.5$ and
 $L=16 \sim 80$. Points denote the QMC data for $J'=0.3$, 0.25, 0.2,
 and 0.15 from top. The lines show the fitting function
 $\Delta^{(2)}(x)$. In the inset the $J'^2$ dependence of the spin gap is
 shown. The line with the slope of 1.5 is the guide to eyes.}
\end{center}
\vspace*{-1ex}
\end{figure}

Let us assume that, in the low-concentration limit ($x \rightarrow 0$),
there exists a mid-gap $\Delta$ induced by dilution inside the original spin
gap, $\Delta_{\rm p}$, of the pure system. The former excitation mode
contributes to $C(\tau)$ in proportion to $x$ so that it may less contribute
to $\Delta^{(2)}$ as compared to the back ground mode of $\Delta_{\rm p}$.
More explicitly, $C(\tau)$ in this situation and in the limit $L \rightarrow
\infty$ may be written as~\cite{todo}
\begin{equation}
 \label{corr}
 C(\tau)=(1-ax)({\rm sinh}\frac{\beta\Delta_{\rm p}}{2})^{-1}{\rm cosh}[(\tau-\frac{\beta}{2})\Delta_{\rm p}] + ax({\rm sinh}\frac{\beta\Delta}{2})^{-1}{\rm cosh}[(\tau-\frac{\beta}{2})\Delta] \ ,
\end{equation}
where $a$ is a non-trivial constant depending on $\alpha$ and $J'$.
Substituting Eq. (\ref{corr}) for Eq. (\ref{delta2}), we obtain
the following relation between $\Delta^{(2)}$, $\Delta_{\rm p}$ and $\Delta$:
\begin{equation}
  \label{fit}
  \Delta^{(2)}(x)=\Delta \sqrt{\frac{1+\frac{1-ax}{ax}\frac{\Delta}{\Delta_{\rm p}}}{1+\frac{1-ax}{ax}(\frac{\Delta}{\Delta_{\rm p}})^3}} \ .
\end{equation}
To confirm that the above assumption is in fact the case, we simulate
the systems in which only one strong bond is removed, i.e., $x=1/2L^2$.
In Fig. 2 the $x$ dependences of the QMC data of $\Delta^{(2)}$ are shown
for $J'=0.3$, 0.25, 0.2, 0.15 and $\alpha=0.5$. They are
well fitted to Eq. (\ref{fit}), the lines in the figure. This implies that
the spin gap $\Delta$, or the mid-gap, is independent of $x$ at least in the
$x$-range shown in Fig. 2. By this fitting the value of $\Delta$ is obtained
as shown in the inset of Fig. 2, in particular $\Delta=0.138(5)$ for
$J'=0.3$ as indicated in Fig. 3. The existence of the spin gap
proportional to $J'^2$ is interpreted that two edge spins induced at
both ends of a diluted bond reform a singlet pair due to the effective
coupling $\tilde{J}_{\rm af}$ ($\sim J'^2$) through the 2D shortest
paths.~\cite{yasuda2} Since two edge spins on the diluted bond are on
the different sublattices, the interaction is always AF.

\begin{figure}[t]
\begin{center}
 \leavevmode
 \epsfxsize=0.6\textwidth
 \epsfbox{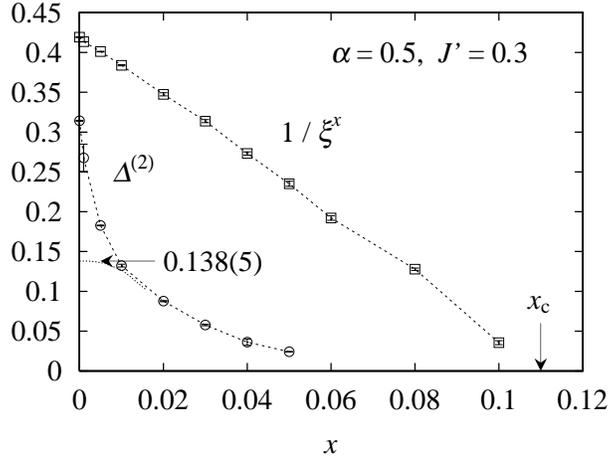}
 \caption{The concentration dependences of $\Delta^{(2)}$ corresponding to
 the upper bound of the spin gap and the inverse spatial correlation
 length for $\alpha=0.5$ and $J'=0.3$. This
 system has the critical concentration $x_{\rm c}=0.11(1)$. The value of
 $\Delta^{(2)}$ at $x=0$ equals 0.31415(2). The dotted
 line describes the spin gap independent of $x$ in the
 extremely-low-concentration region and the expected reduction in the
 low-concentration region.
 }
\end{center}
\vspace*{-1ex}
\end{figure}

As $x$ increases, the distinction between a mid-gap and the gap
$\Delta_{\rm p}$ becomes obscure. Instead, the value of $\Delta^{(2)}$ gives
us an upper bound of the spin gap $\Delta$ of our interest, whose
$x$-dependence is schematically shown by the dotted line in Fig. 3.
Presumably $\Delta^{(2)}$ is a good estimate for $\Delta$ in the range $x
\nge 0.02$.

In Fig. 3 we also show the $x$ dependence of the spatial correlation length
$\xi^x(x)=\lim_{L \rightarrow \infty}\xi^x(L;x)$ at $T \rightarrow 0$. The critical concentration
$x_{\rm c}$ is about 0.11, which is estimated by the finite-size-scaling
analysis using the QMC data of $\xi^x(L;x)$. Its details will be reported
elsewhere. In marked contrast to $1/\xi^x$, $\Delta^{(2)}\ (\simeq \Delta$
in the range $x\nge 0.02$) seems to vanish at a certain value $x'_{\rm c}$
smaller than $x_{\rm c}$, which strongly suggests the existence of the
quantum Griffiths phase at $x'_{\rm c}<x<x_{\rm c}$. Thus the bond-diluted
system has the more peculiar properties than the site-diluted system.

The $x$ dependence of the N\'{e}el temperature in a bond-random
system CuGe$_{1-x}$Si$_x$O$_3$ has been
investigated experimentally,~\cite{regnault,masuda2} and has turned out
to be similar to that of the site-diluted system
Cu$_{1-x}$Mg$_x$GeO$_3$. Besides the problem whether the
substitution of Si for Ge simply corresponds to the bond dilution or not,
we have to investigate the systems at finite but low temperatures,
explicitly taking into account the interlayer coupling. Experimentally,
analyses on systems with further smaller $x$ and at lower temperatures
are desirable.

\section{Summary}

We investigated dilution effects on the spin-1/2 quantum AF Heisenberg
model with the spin-gapped ground state. When the strong bonds are
removed from the lattice for the spin-gapped ground state, two effective
interactions with the couplings $\tilde{J}_{mn}$ and $\tilde{J}_{\rm
af}$ are induced between the effective spins. The competition between
$\tilde{J}_{mn}$ and $\tilde{J}_{\rm af}$ leads to peculiar quantum
phase transitions. In the low-concentration region, $\tilde{J}_{\rm af}$
between edge spins mainly contributes to magnetic properties and,
therefore, the new spin gap due to singlet pairs reformed by
$\tilde{J}_{\rm af}$ is induced within the original spin
gap. The value of the spin gap decreases as $x$ increases and tends to vanish
at $x'_{\rm c}$ which is less than $x_{\rm c}$ or just at $x_{\rm
c}$. If the former is the case,
the quantum Griffiths phase would exist. Thus there are quite different
magnetic properties at $T=0$ between the site- and bond-diluted systems. 

\section*{Acknowledgements}
Most of numerical calculations in the present work have been performed
on the SGI 2800 at Institute for Solid State Physics, University of
Tokyo. The program is based on 'Looper version 2' developed by S.T. and
K. Kato and 'PARAPACK version 2' by S.T. The present work is supported
by the ``Research for the Future Program'' (JSPS-RFTF97P01103) of Japan
Society for the Promotion of Science. S.T. acknowledges support of the
Swiss National Science Foundation.



\end{document}